\documentclass[conference]{IEEEtran}
\usepackage{cite}
\usepackage{stackengine} 
\newcommand\oast{\stackMath\mathbin{\stackinset{c}{0ex}{c}{0ex}{\ast}{\bigcirc}}}
\usepackage{mathtools}
\usepackage{graphicx}
\graphicspath{ {images/} }
\usepackage{textgreek}
\usepackage{cleveref}
\usepackage{subcaption}
\usepackage{xcolor}
\usepackage{tabularx}
\usepackage{multirow}
\usepackage{textcomp}
\usepackage{array}
\usepackage{amsmath}
\usepackage{amsfonts}
 \IEEEoverridecommandlockouts
 \makeatletter
\newcommand{\printfnsymbol}[1]{%
  \textsuperscript{\@fnsymbol{#1}}%
}
\makeatother
\begin{document}


\title{Multi-Class Micro-CT Image Segmentation Using Sparse Regularized Deep Networks  }


\author{Amirsaeed Yazdani\textsuperscript{*}$^{1}$\thanks{\textsuperscript{*}These authors contributed equally}, Yung-Chen Sun\printfnsymbol{1}$^{1}$, Nicholas B. Stephens$^{2}$, Timothy Ryan$^{2}$, Vishal Monga$^{1}$\\ $^{1}$ \small{Department of Electrical Engineering,} $^{2}$ \small{Department of Anthropology, The Pennsylvania State University, University Park, Pennsylvania, USA.}   }
\maketitle

%

\begin{abstract}
It is common in anthropology and paleontology to address questions about extant and extinct species through the quantification of osteological features observable in micro-computed tomographic (\textmu CT) scans. In cases where remains were buried, the grey values present in these scans may be classified as belonging to air, dirt, or bone. While various intensity-based methods have been proposed to segment scans into these classes, it is often the case that intensity values for dirt and bone are nearly indistinguishable. In these instances, scientists resort to laborious manual segmentation, which does not scale well in practice when a large number of scans are to be analyzed. Here we present a new domain-enriched network for three-class image segmentation, which utilizes the domain knowledge of experts familiar with manually segmenting bone and dirt structures. More precisely, our novel structure consists of two components: 1) a representation network trained on special samples based on newly designed custom loss terms, which extracts discriminative bone and dirt features, 2) and a segmentation network that leverages these extracted discriminative features. These two parts are jointly trained in order to optimize the segmentation performance. A comparison of our network to that of the current state-of-the-art U-NETs demonstrates the benefits of our proposal, particularly when the number of labeled training images are limited, which is invariably the case for \textmu CT segmentation.
\end{abstract}

\begin {IEEEkeywords}
Micro-CT, Osteology, Fossil, Quantitative morphometrics, Deep learning, Neural networks, U-NET
\end{IEEEkeywords}

\section{Introduction}
\label{intro}
\begin{figure}[t!]
         \centering
          \begin{subfigure}[t]{0.4\columnwidth}
        \centering
        \includegraphics[height=3cm]{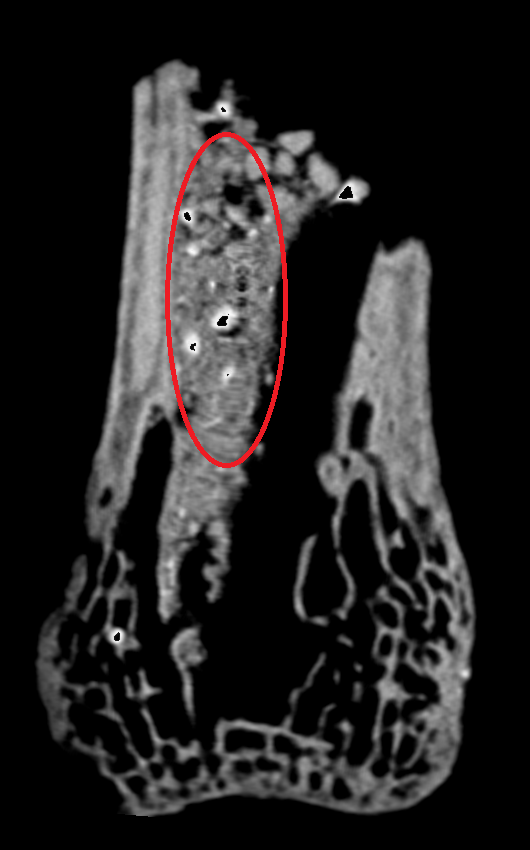}
        \caption{Original}
    \end{subfigure}%
        \begin{subfigure}[t]{0.4\columnwidth}
        \centering
        \includegraphics[height=3cm]{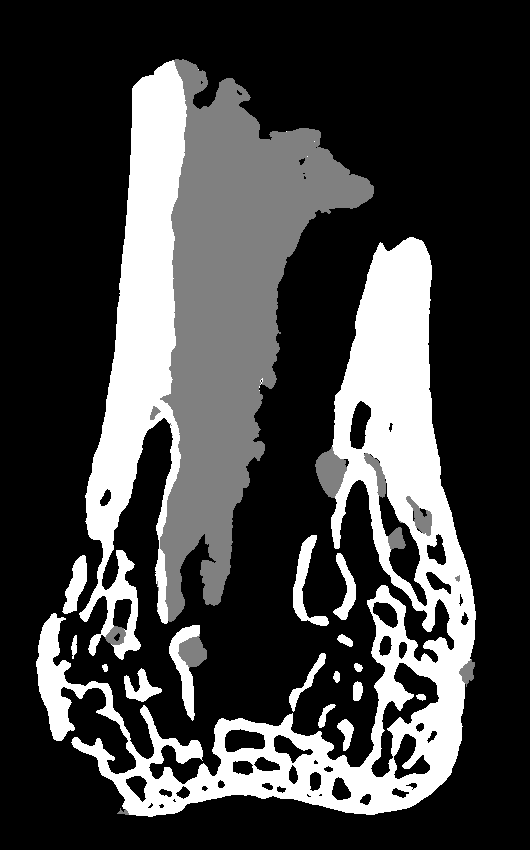}
        \caption{Ground truth}
    \end{subfigure}%
        \caption{ A sample of an unsegmented image and its ground truth. The region inside the red circle looks like bone, however it is actually dirt (gray pixels in the ground truth)}
        \label{fig:Original}
 \end{figure}

Micro-computed tomography (\textmu CT) is increasingly used in palaeontology and anthropology to address fundamental questions about functional morphology, evolutionary history, and phylogenetic constraint \cite{Witmera,Brassey2012,Skinner2010,Harvati2019}. Often these analyses rely on the quantitative analysis of bone, which is frequently the only remaining tissue that survives long-term burial after an animal dies. As such, their accuracy relies on a faithful segmentation of \textmu CT scan intensity values belonging to air, bone, and non-bone material. In burial contexts, non-bone intensity values tend to belong to extraneous soils of varying densities, which are introduced following soft tissue decomposition. While the intensity values belonging to soil may differ dramatically from those of bone and air, there are many cases where heterogeneous densities form a continuum of values between bone and non-bone pixels, which limits the efficacy of intensity-based segmentation approaches\cite{Doube2010,Otsu1979,Odgaard1993,Ridler1978}. In these instances, laborious manual segmentation is performed by an expert prior to analysis. An example of an unsegmented image and its corresponding ground truth can be seen in figure\ref{fig:Original}.

While slice-by-slice manual segmentation is able to resolve many ambiguities that arise, it is untenable when a single high-resolution scan may result in a 3D volume with large grids (2040 x 2040) and thousands of tomographic slices. Sophisticated automatic approaches have been proposed, such as the k-means with fuzzy c-mean clustering algorithm \cite{Dunmore2018}, but intensity-based clustering results in occasional misclassification of non-bone material adjacent to bone. Alternative edge detection methods have also been applied to palaeontological material\cite{Scherf2009}, but the implementation is not freely available. There are very few examples of machine learning applications to the segmentation of trabecular bone structure \cite{MLP}, which are limited by the quantity and quality of training data. 

 \begin{figure*}
         \centering
         \begin{subfigure}[t]{\textwidth}
        \centering
         \includegraphics[height=2cm]{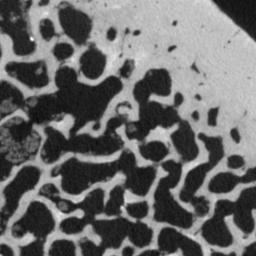}
         \includegraphics[height=2cm]{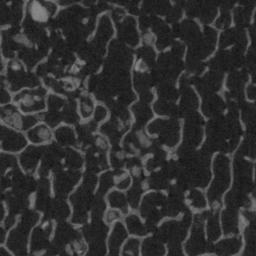}
         \includegraphics[height=2cm]{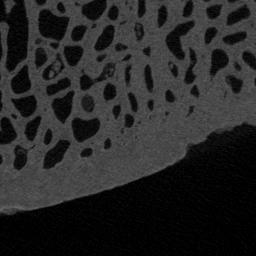}
         \includegraphics[height=2cm]{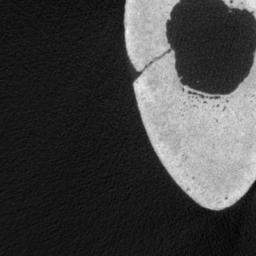}
         \includegraphics[height=2cm]{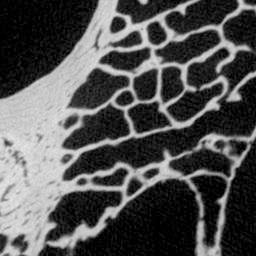}
         \includegraphics[height=2cm]{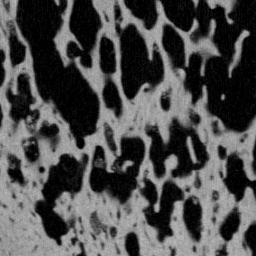}
         \includegraphics[height=2cm]{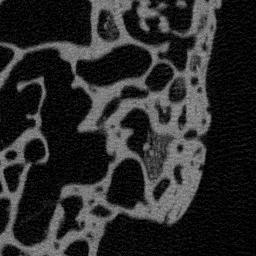}
         \includegraphics[height=2cm]{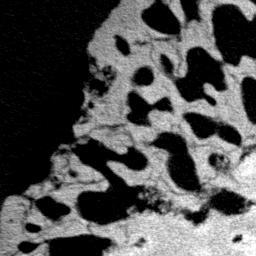}
        \caption{}
    \end{subfigure}
    \newline
       \begin{subfigure}[t]{\textwidth}
        \centering
       \includegraphics[height=2cm]{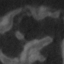}
        \includegraphics[height=2cm]{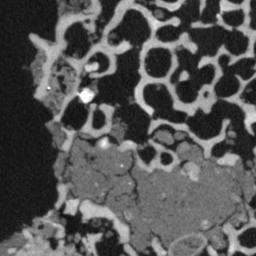}
        \includegraphics[height=2cm]{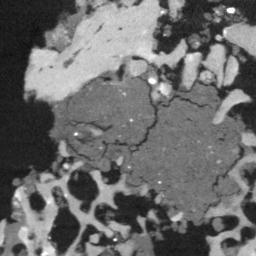}
        \includegraphics[height=2cm]{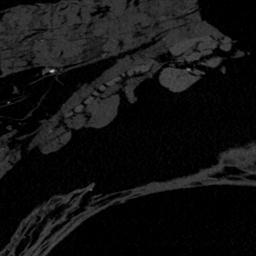}
        \includegraphics[height=2cm]{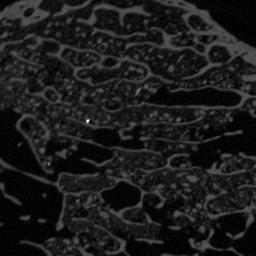}
        \includegraphics[height=2cm]{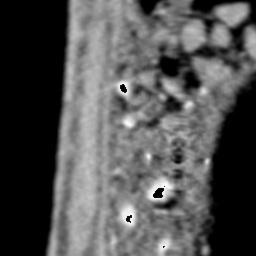}
        \includegraphics[height=2cm]{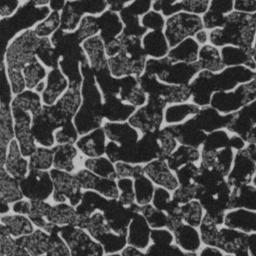}
        \includegraphics[height=2cm]{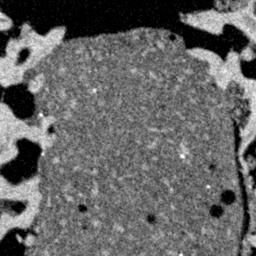}

        \caption{}
    \end{subfigure}%
          
        \caption{ Examples of bone (a) and dirt (b) patches. Although in terms of pixel intensity the two groups look very similar, they have some discriminative features that the network tries to capture.  }
        \label{fig:visual 2}
 \end{figure*}

Deep learning frameworks applied to various imaging tasks have shown promising results, with image segmentation problems being of particular note \cite{18,19,20,21,22,23,24,U-net,asilomar}. In these cases, correspondence between unsegmented images and their segmented labels are mapped via sets of non-linear activation functions. Despite their success in segmentation tasks, their accuracy usually degrades when salient features are lacking in the underlying raw data. Solving this problem requires the incorporation of domain knowledge into the training dataset for the segmentation network to succeed. Previously, \cite{asilomar,Cherukuri,zoori,ajpa} developed networks that exploit special characteristics of segmentation problems using geometric priors. Even so, their prior guided segments only two classes, which for our problem would entail combining dirt and air into a general non-bone class. As depicted in figure 1, the similarities between bone and dirt intensity values heighten the risk of erroneously classifying soil as bone. However, despite the similarities between these two classes, they differ in terms of connectivity and distribution over all structures. In this work we propose a novel regularized network for three-class segmentation of bone, which consists of 1) custom representation layers where discriminative features of bone and dirt are extracted and 2) a segmentation network which uses the extracted features from the representation layers to segment the images. The representation layers have two blocks, one for each class, which are themselves trained using representative dirt and bone patches extracted from the dataset using custom regularization terms in addition to the segmentation loss. In particular, the regularizer for the 'dirt' class explicitly enforces sparsity, capturing the relatively small spatial footprint of dirt structures in the images. These patches are chosen such that the bone block has a high response to bone patches and a low response to dirt patches, and vice versa. We refer to this proposed network as a "Discriminative Sparse Regularized Deep Network for \textmu-CT segmentation" (DS-RDN).

\section{Discriminative Sparse Regularized Deep Networks for \textmu CT Segmentation}
\label{Methodology}
\subsection{Background and Notation}
Let $X \in \mathbf{R}^{M\times N}$ represent the input image. Let $Y\in \mathbb{R}^{M\times N}$ be the output segmented image and $Y_{g} \in \{0,1,2\}^{M\times N}$ be the manually-labeled y segmentation map corresponding to $X$ where 0, 1, and 2 represent air, dirt, and bone, respectively. Note that this type of labeling essentially means that the outputs/labels for the network are vectors/matrices with three channels, in which each channel corresponds to each class and is set to one when a certain pixel belongs to that class (one-hot encoding). Let $W_{R_{k}}^{B}$ and  $W_{R_{k}}^{D} \in \mathbb{R}^{m\times n}$ be the $k^{th}$ filter in the geometric representation layer of bone and dirt, respectively, where $m$, $n$ represent the width and height of the filter, respectively. Similarly, let $W_{S_{k}}^{l} \in \mathbb{R}^{m\times n\times d}$ be the $k^{th}$ convolutional filter in layer $l$ of the segmentation network, where $m$, $n$, and $d$ represent the width, height, and depth of the filter, respectively. The representation and segmentation networks are, $\Theta_{R} = \{W_{R_{k}}^{B},W_{R_{k}}^{D}\} \forall k$ and $\Theta_{S} = \{W_{S_{k}}^{l}\} \forall l,k$, respectively. Finally, let the mapping function of the representation network be represented by $f$, and the mapping function of the segmentation network be represented by $F$. Then, $Y = F(f(X, \Theta_{R}), \Theta_S)$. The network parameters $\Theta_{R}$ and $\Theta_{S}$ are learned by minimizing a loss function so as to produce an output $Y$ that closely mimics the ground truth map $Y_{g}$. Since we are dealing with multi-class segmentation we use the cross entropy loss function where we have:
	\begin{equation}
	\label{eq:reg}
	L(\Theta_{S}, \Theta_{R}) = -\sum_{i=1}^{3}Y_{g}^{i}\log \Tilde{(Y^{i}})
	\end{equation}
	Where $Y_{g}^{i}$  is the i-th channel of ground truth and $\Tilde{Y^{i}}$ is the network output after being passed to the Sigmoid function. From a probabilistic point of view, we aim to make the distribution of the output as close as possible to the ground truth distribution for each class. 


\subsection{Representation Layer}The representation layers are trained jointly with the segmentation network to obtain two blocks specifically optimized for catching bone or dirt features.
Feature extraction is complicated as a result of the shared intensity values and organization of dirt and bone (figure\ref{fig:Original}). Therefore we define two blocks, each of which is in charge of detecting bone or dirt features. We define bone-dirt constraints for these blocks to grasp discriminative features and discard similarities. These constraints are imposed on the representation layer filters but they indirectly influence the segmentation network, since their joint optimization with the segmentation network propagates the changes.

\subsection{Bone-Dirt Constraint}
To obtain bone/dirt features, each of the bone/dirt blocks should be sensitive to bone/dirt patches and non-sensitive to dirt/bone patches. To achieve this, we define loss terms for each of the blocks so that they will be optimized with regard to their dedicated class features. Extractions of bone/dirt patches were carried out on challenging portions considered to be faithful representatives of each  corresponding class. Figure \ref{fig:visual 2} shows some of these patches for each class, where the bone patches in figure \ref{fig:visual 2} a) and dirt patches in figure \ref{fig:visual 2} b) are very similar in terms of pixel intensity. Optimization of the bone/dirt blocks is achieved by defining:

\begin{equation}
\tiny
    L_{Bone}(\theta_{R})=\sum_{n=1}^{K}\sum_{i=1}^{P}(\alpha||W_{R_{n}}^{Bone}\oast I_{d}^{i}||_{2}^{2}-\beta||W_{R_{n}}^{Bone}\oast I_{b}^{i}||_{2}^{2})
\end{equation}
\begin{equation}
\tiny
    L_{Dirt}(\theta_{R})=\sum_{n=1}^{K}\sum_{i=1}^{P}(\gamma||W_{R_{n}}^{Dirt}\oast I_{b}^{i}||_{2}^{2}-\sigma||W_{R_{n}}^{Dirt}\oast I_{d}^{i}||_{2}^{2} \\
    +\zeta||W_{R_{n}}^{Dirt}\oast I_{d}^{i}||_{1})
\end{equation}
\normalsize
Where $I_{d}^{i}$ and $I_{b}^{i}$ are the $i^{th}$ dirt and bone patches, respectively, $K$ is the total number of representation layer filters, with $\alpha$, $\beta$, $\gamma$, $\sigma$, and $\zeta$ being positive regularization constants. The negative signs mean our goal is to maximize those terms while positive signs mean minimization objectives. We also add a sparsity term to aid in extracting features from the dirt block, because structures in the dirt class patterning occupy a relatively small part of the image, hence making it sparse.

     \begin{figure*}
         \centering
        \includegraphics[width=.75 \textwidth]{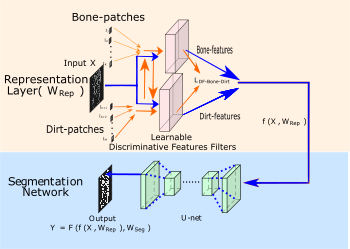}
        \caption{ The structure of the proposed network. The representation layer consists of two blocks, one each for bone and dirt. These blocks are trained jointly using the normal training images and extracted patches containing specific patterns, which are then used to compute the bone and dirt loss ($\boldsymbol{L_{Bone}}$ and $\boldsymbol{L_{Dirt}}$ ) }
        \label{fig:network}
 \end{figure*}
 
 \subsection{Segmentation Network}
 The features extracted by the representation layer are concatenated and passed to a modified U-NET segmentation network \cite{U-net}, which is an architecture that has been shown to perform well in various classification and segmentation problems \cite{asilomar,23}. The U-NET modifications are as follows: 1) the input channels are adjusted based on the number of channels from the representation layers and 2) the number of output channels is increased to three for generating one-hot outputs. So the complete loss function of the domain-enriched network would be:
 \begin{equation}
     L(\Theta_{S}, \Theta_{R}) = -\sum_{i=1}^{3}Y_{g}^{i}\log \Tilde{(Y^{i}})+\lambda_{1}L_{Bone}+\lambda_{2}L_{Dirt}
 \end{equation}
 where $\lambda_{1}$ and $\lambda_{2}$ are regularizer constants. The parameters of the representation layer ($\theta_{R}$) and segmentation network ($\theta_{S}$) are optimized using the stochastic gradient descent \cite{lecun1998gradient,werbos1994roots}. Since all of the terms in the loss function are differentiable with regard to network parameters, standard back-propagation is performed. The proposed domain-enriched network is illustrated in figure \ref{fig:network}:

\section{Experimental Evaluation}
\label{Experimental Evaluation}
\noindent\textbf{{Datasets and Experimental Setup}}: Human, non-human, and fossil post-cranial osteological samples (n$\simeq$30) were \textmu CT scanned (30-50\textmu m voxel resolution) either at the Center for Quantitative Imaging, Pennsylvania State University using an ONMI-X HD600 or GE v\textbar tome\textbar x L300 (180 kV, 110 mA, 2800-4800 views), or at the Cambridge Biotomography Centre, University of Cambridge using a Nikon XTH 225 ST HRCT (125 kV, 135 mA, 1080 views). Ground truth images were hand-segmented with a digital tablet by an expert with n$\simeq$9 years experience. Networks were trained with 20 images and evaluated using a set of 13 images. Two additional experiments were performed to 1) evaluate the performance based on a lower training set of 10 images and b) to evaluate the robustness of the networks by randomly choosing the train/test splits.  \\
\noindent \textbf{Network Architecture:} The domain bone/dirt blocks are composed of $K=16$ $3\times 3$ filters using a Light-UNet segmentation network as proposed in, which is obtained by compressing the U-NET layers \cite{U-net} 

\begin{figure}[h]
\centering
\includegraphics[height=2.5cm]{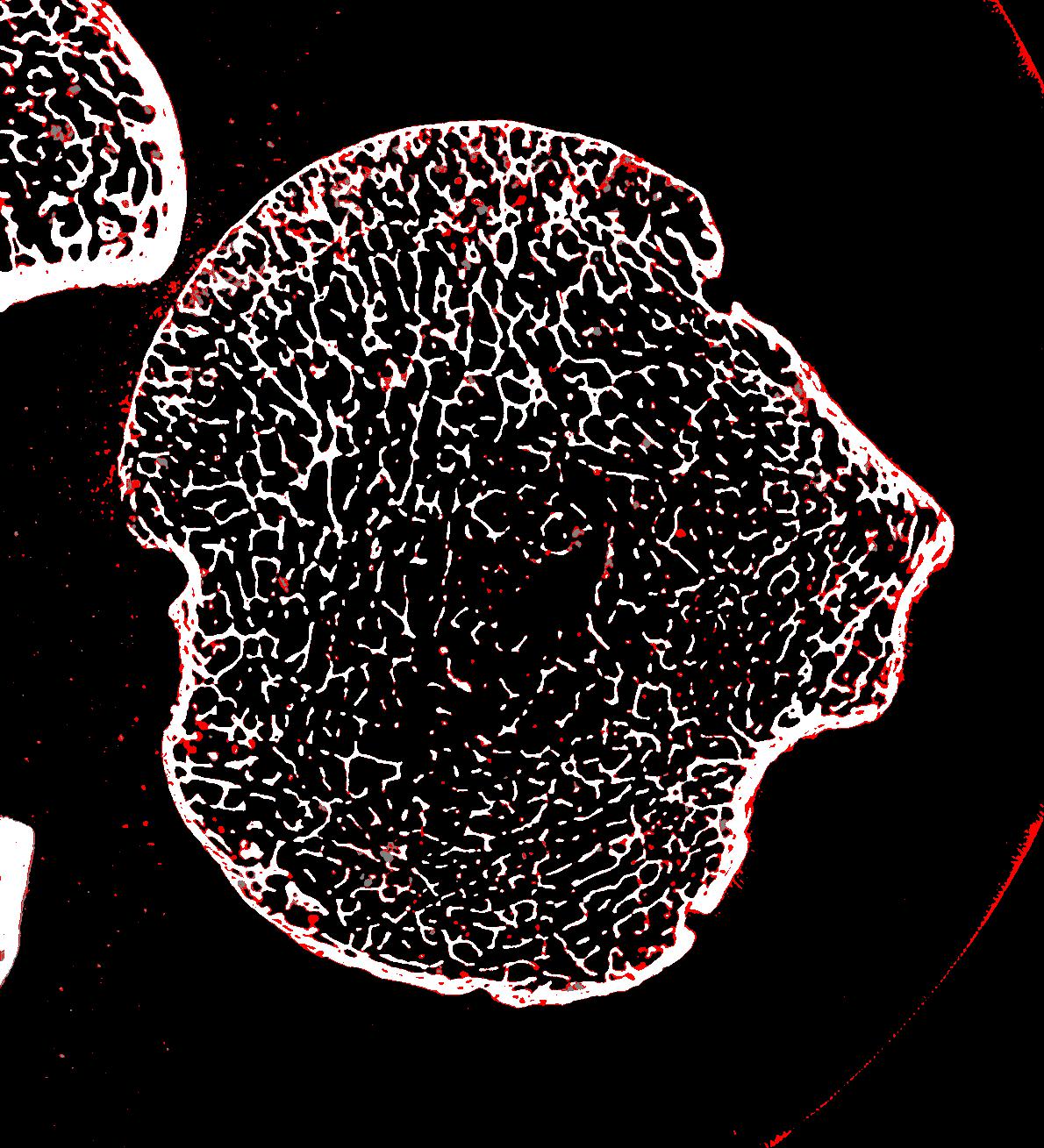}
\includegraphics[height=2.5cm]{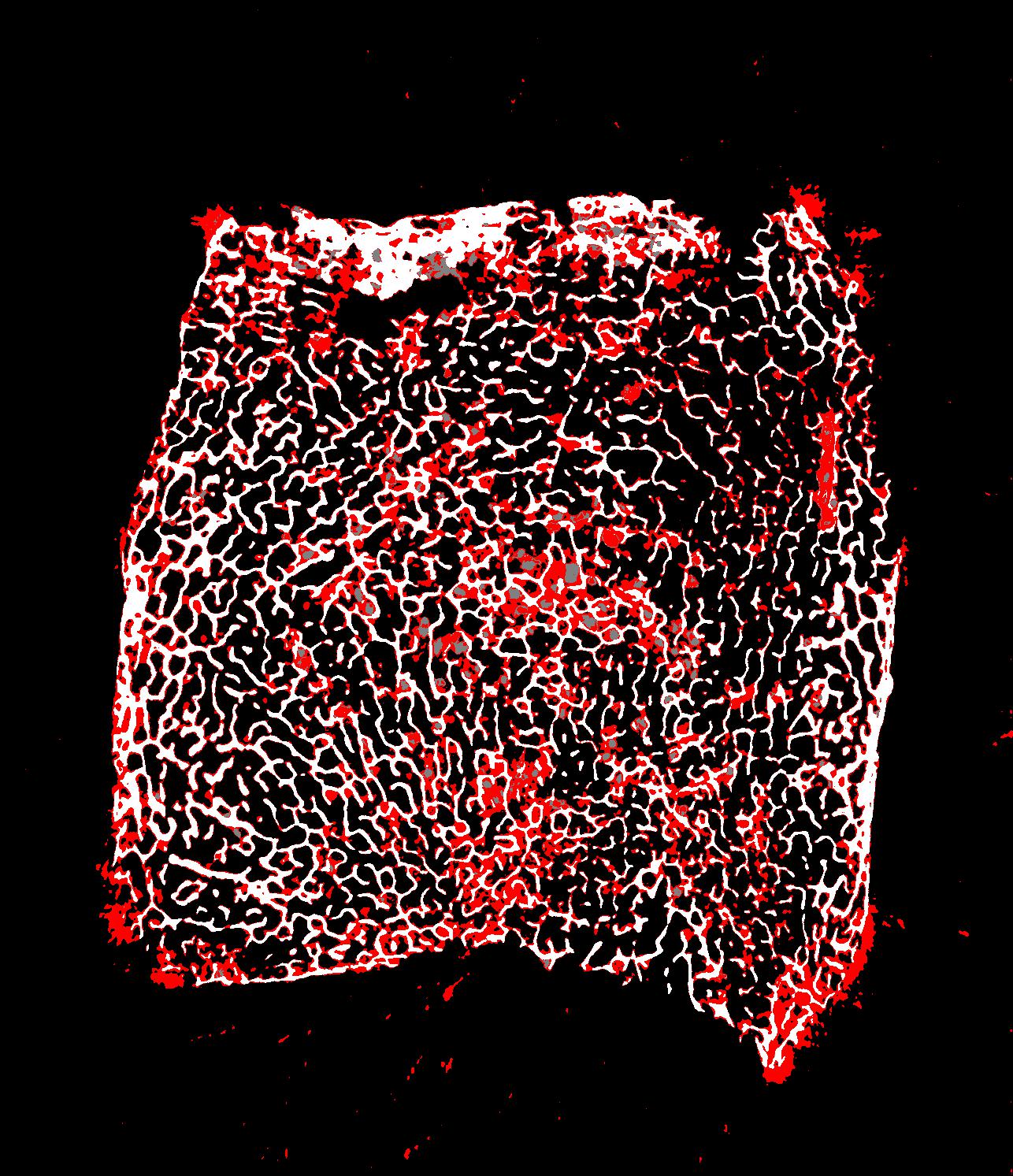}\\
\includegraphics[height=2.5cm]{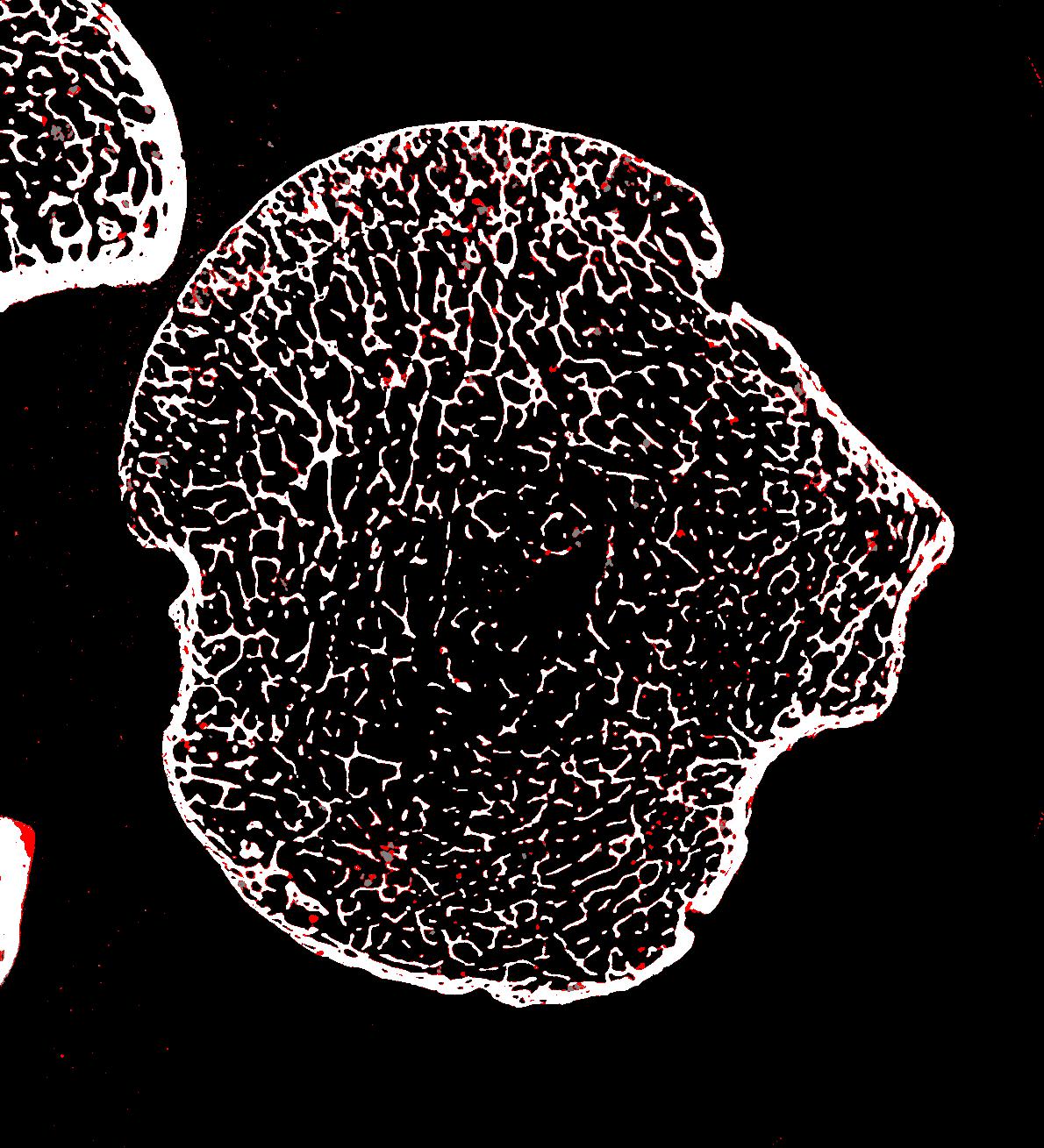}
\includegraphics[height=2.5cm]{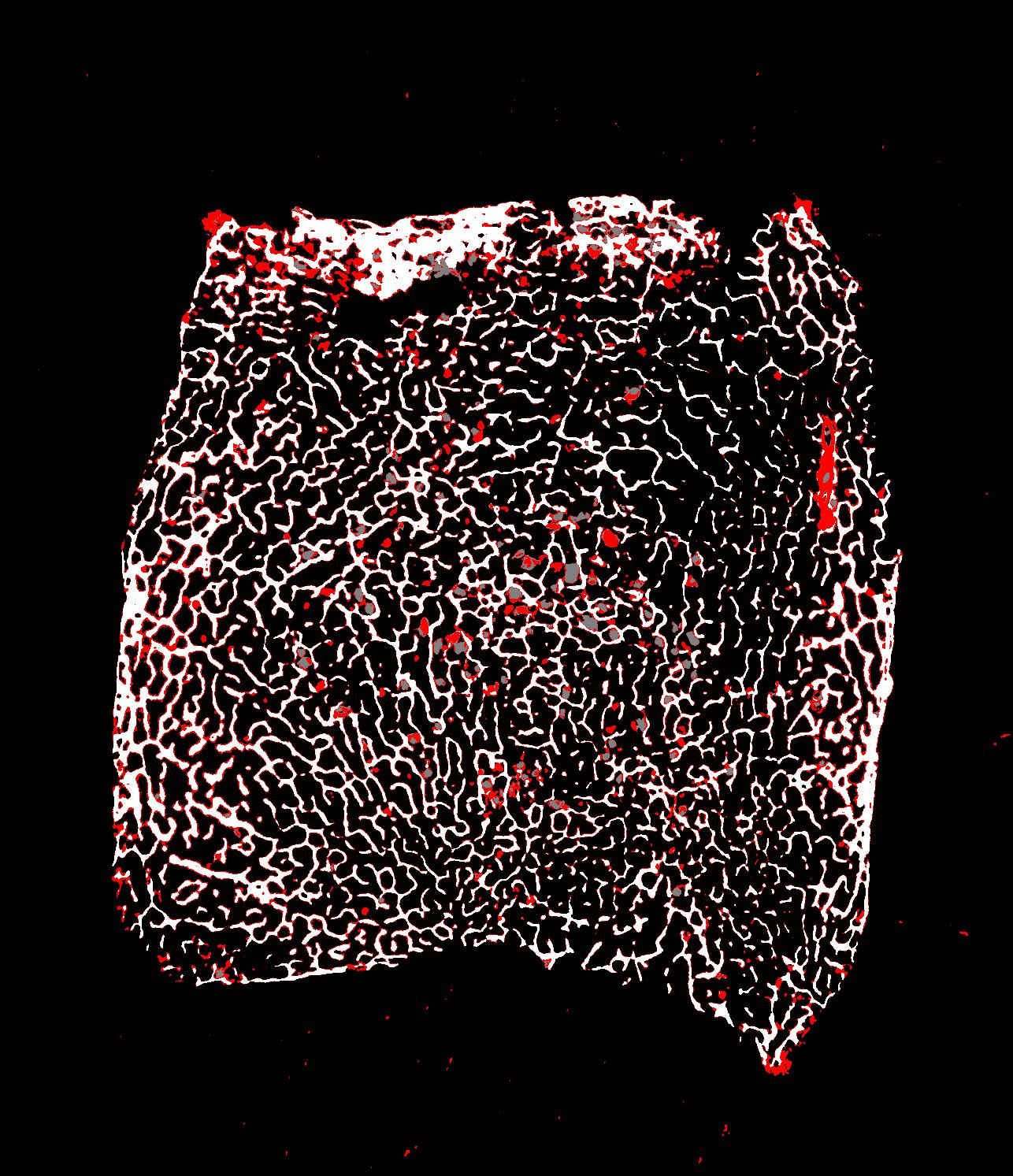}
        \caption{The comparison of errors in two test samples segmented by a typical U-NET (top) and DS-RDN (bottom). Red pixels represent errors in segmentation across all classes.}
        \label{fig:error}
    \end{figure}

\begin{figure}
\centering
\includegraphics[height=3cm,width=9cm]{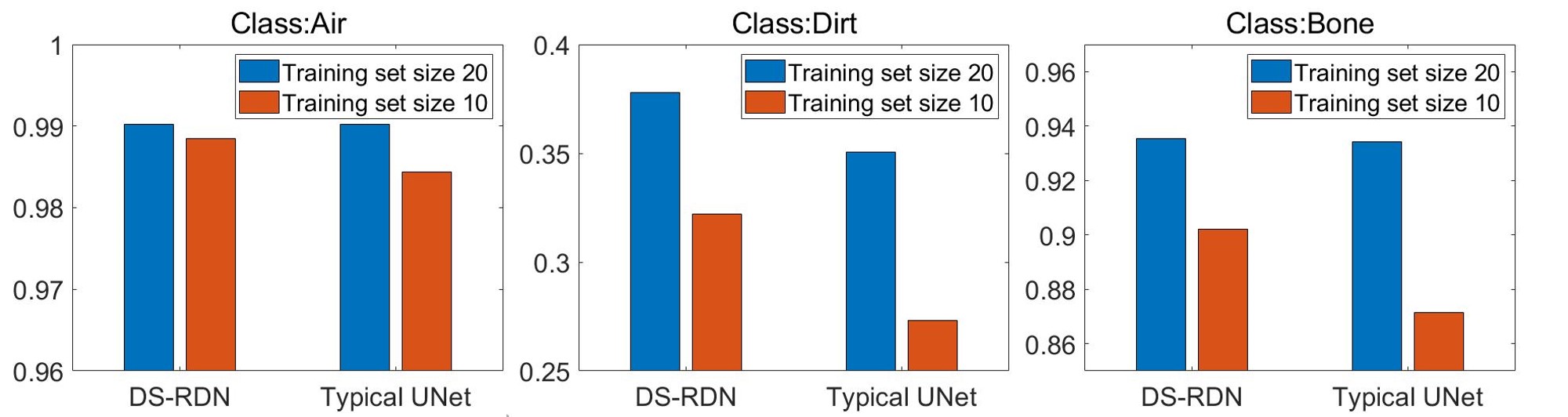}
        \caption{Change in the Dice overlap in the first experiment as the training size decreases from 20 to 10 for DS-RDN and typical U-NET. From left to right: air, dirt and bone.}
        \label{fig:experiment1}
    \end{figure}

\begin{figure}
\centering
\includegraphics[height=3cm,width=9cm]{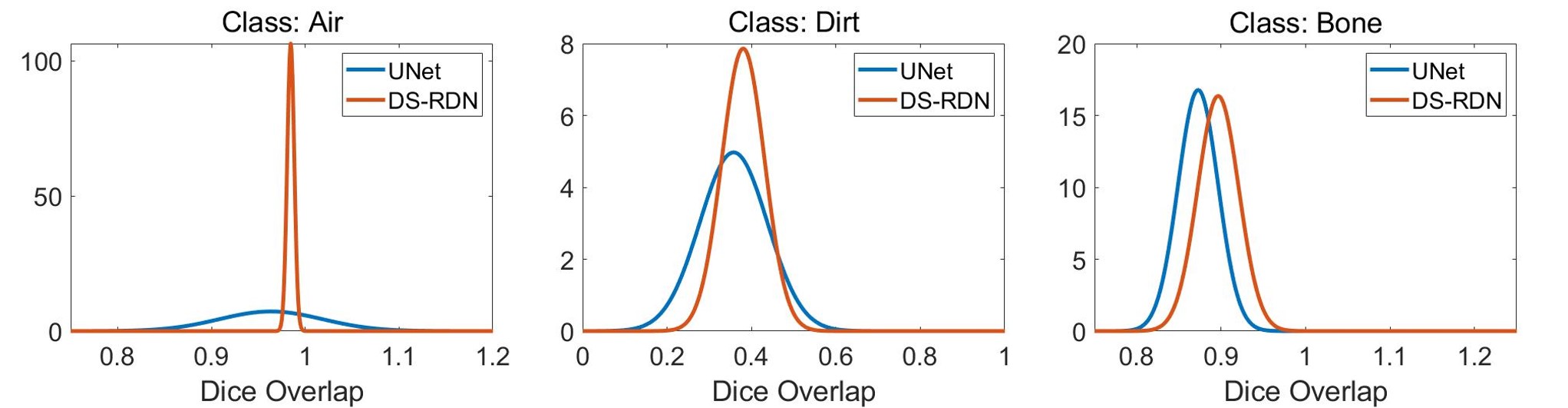}
        \caption{ The Gaussian distributions obtained by taking the mean and the variance of 10 different train/test splits (second experiment). From left to right: air, dirt, bone.}
        \label{fig:experiment2}
    \end{figure}

\noindent\textbf{Metrics for Evaluation}:
Dice overlap (F1) is used as an evaluation metric for each class individually, with the definitions $F1 = \frac{2TP}{2TP+FP+FN}$, with $TP$, $FP$, and $FN$ representing true positives,  false positives, and false negatives, respectively with regard to each certain class.\\ 
\noindent\textbf{Patch Extraction and Parameter Selection}: Following from other deep learning frameworks \cite{18, 19, U-net}, nearly 1200 $256\times 256$ patches were extracted from the original images by using sliding windows stride length set at 32. 
In any given dataset, the number of bone patches might not equal the number of dirt patches. Therefore, in every training epoch, we randomly selected an equal number of both types of patches. The regularization parameters inside the representation loss were 1e-5 and 0.5 for $\alpha$, $\beta$, $\lambda$, and $\eta$, respectively.
\noindent The batch size was set to 44 with 25 training epochs. The learning rate started at $10^{-4}$ and decreased with a drop factor of 0.1 for every 8 epochs. Training and testing were conducted on an NVIDIA Titan X GPU (12GB) with the PyTorch package \cite{NEURIPS2019_9015}.\\
\noindent\textbf{Normal Set-up}:Table \ref{tab:table} shows a comparison of Dice overlap results for the full training set of 20 images and 13 evaluation images for the proposed network (RDN-CS), a typical U-NET, and MLP. The higher value (specifically for the dirt class) shows the efficacy of including bone/dirt blocks in our method.\\
Figure\ref{fig:error} visualizes the differences in segmentation accuracy in two images using the proposed network and a typical U-NET. The incorrectly classified pixels across all classes are shown in red, with the proposed network having fewer incorrectly classified pixels in both images.
    \begin{table}
    \centering
    \caption{Comparison between different methods }
    \begin{tabular}{|c|c|c|c|}
        \hline
                 \multirow{2}{*}{Method} &\multicolumn{3}{c|}{Dice Overlap} \\\cline{2-4}
         &Air&Dirt&Bone\\\hline
         MLP \cite{MLP} &0.9890 & 0.2193&0.8526 \\\hline
         U-net \cite{U-net} &0.9902 &0.3507&.9342 \\\hline
         RDN-CS  & 0.9902 & \textbf{0.3780}&\textbf{0.9354}\\\hline
    \end{tabular}
    
    \label{tab:table}
\end{table}
\\\noindent\textbf{First Experiment}: To determine how incorporating domain knowledge into our network affected performance at lower training sizes, we trained our network and a typical U-NET using 10, instead of 20, training images. Figure \ref{fig:experiment1} shows the Dice overlap results for each class using the original evaluation set. These results demonstrate that the proposed method performs well despite the halving of the training sample. Specifically, while there is a substantial degradation of performance in the typical U-NET (especially for the dirt class), performance degradation of the DS-RDN is relatively minimal.
\\\noindent\textbf{Second Experiment}: In order to assess how robust the networks were regarding variation in the train/test splits, we performed 10 separate training sessions of our network and a typical U-NET, using 10 different train/test splits of 20 images. The results depicted in Figure \ref{fig:experiment2} show the Gaussian distribution obtained by considering the mean and variance of the Dice overlap over all splits for each architecture, with respect to each class. When compared to the distribution of the typical U-NET, the higher mean and narrow distribution of the proposed network demonstrates that it is more robust against variation in the train/test split for each class.




    
 


\section{Conclusion}
\label {Conclusion}

Here we proposed a regularized custom network for multi-class segmentation of bone structure that incorporated domain knowledge to discriminate between two challenging classes. Multiple experimental designs were considered to measure the performance and stability of the proposed architecture versus well-established alternatives. When the training size was halved, we found the degradation of the proposed network to be minimal compared to a typical U-NET. Similarly, we observed higher mean and narrower standard deviation distributions when varying the train/test data, indicating that the proposed network is more robust than a typical U-NET. Ultimately, these results support the incorporation of domain-specific knowledge in challenging segmentation tasks.


\bibliographystyle{IEEEtran}
\bibliography{main.bib}

\end{document}